\documentclass{article}
\usepackage{arxiv}
\usepackage[utf8]{inputenc} % allow utf-8 input
\usepackage[T1]{fontenc}    % use 8-bit T1 fonts
\usepackage{hyperref}       % hyperlinks
\usepackage{url}            % simple URL typesetting
\usepackage{booktabs}       % professional-quality tables
\usepackage{amsfonts}       % blackboard math symbols
\usepackage{nicefrac}       % compact symbols for 1/2, etc.
\usepackage{microtype}      % microtypography

\usepackage{graphicx}
\usepackage{doi}
\usepackage{setspace}
\usepackage{mathtools}
\usepackage{wasysym}
\usepackage{latexsym}

\title{Generation of 1 Gb full entropy random numbers with the enhanced-NRBG method}

\date{} 					% Or removing it
        
\author{ {Deepika   Aggarwal} 
	
	\And
	{Karthick Balaji R} 
	\And
	{Rohit Ghatikar} 
	\And
	{Sruthi Chennuri  } 
	\And
	{Anindita Banerjee \thanks{aninditabanerjee.physics@gmail.com}} \\
	 QuNu Labs Pvt. Ltd. \\
              M.G. Road, Bangalore, India\\
	}

\begin{document}
\maketitle

\begin{abstract}
Random numbers have significant applications in fundamental science, high-level scientific research, cryptography, and several other areas where there is a pressing need for high-quality random numbers. We present an experimental demonstration of a non-deterministic random bit generator from a quantum entropy source and a deterministic random bit generator mechanism to provide  high quality random numbers providing a  throughput of 1 Gb. Quantum entropy is realized by a series of quantum chips based on radioactive isotope Americium-241.  The extracted raw random numbers are further post-processed to generate a high-entropy seed for the hash based deterministic random bit generator. We discuss the implementation of randomness extraction algorithm and Hash-DRBG algorithm in detail.  The random numbers pass all randomness measures provided in   ENT and NIST test suites.
\end{abstract}

% keywords can be removed

\section{Introduction}\label{intro}

Random numbers are a sequence of numbers that cannot be reasonably
predicted better than by a random chance. The characteristics of true
randomness are unpredictability, uniform distribution of the bits
in the sequence and lack of patterns in the sequence. Random numbers
are useful for a variety of purposes, such as simulation, cryptography,
authentication, online gaming/lotteries, statistical sampling and
many more. There are two approaches to random number generation: algorithmic and physical. Pseudo-random number generator (PRNG) which is also known as deterministic random bit generator (DRBG) uses mathematical algorithms that produce long sequences of apparently random results, i.e. they appear to be statistically independent and unbiased but are completely determined by a shorter initial value, known as a seed. As a consequence, the entire seemingly random sequence can be reproduced if the seed value is known. The output from the PRNG remains fundamentally non-random and predictable since it is generated by a deterministic algorithm.
Random numbers are repeated after many iterations, and the same seed
always creates the same numbers. On the other hand, a true random number generator (TRNG) extracts randomness from the observation of physical processes that behave in a non-deterministic way. The physical phenomena based on quantum physics are provably random. This makes them better candidates for true random number generation.

In this work, we have used radioactive decay as a source of quantum entropy. In \cite{randy,QEC}, the   emission of alpha particles during the decay of the radioactive isotope (Americium-241) is utilized to generate random numbers.  

This behaviour can be mathematically modelled by the Poisson distribution \cite{review}. The distance between consecutive decay events follows negative exponential distribution.
The fact
that the time interval between successive radioactive decay or the number of alpha particles radiated within a constant time is identically and independently distributed, can be used to extract entropy. 
The extracted entropy can be further digitized to obtain the random bits. These digitized raw random sequences have some non-negligible bias owing to the exponential time distribution between the random events. In order to eliminate the bias we perform post-processing of the raw bit sequence. Post-processing compresses the raw random sequence and generates an output which  is as close as possible to a uniform distribution \cite{review}. We take this refined  sequence as a seed to generate larger throughput. This is achieved by the National Institute of Standards and Technology (NIST)-recommended cryptographic algorithm  for DRBGs such as Hash-DRBG, HMAC-DRBG, and CTR-DRBG \cite{nist90a}. In this scheme, the seed, which is to be fed to the DRBG for its instantiation
is collected from true randomness sources and high quality, unpredictable random numbers with large throughput can be generated. The entire process is termed as enhanced-non deterministic random bit generator (enhanced-NRBG).

This paper is organized as follows. In section \ref{architechture}, we have presented the architecture of the random number generator and in section \ref{QEC}, the quantum source is discussed. Randomness extraction methods and post-processing are explained in section \ref{algo} and \ref{PP}, respectively. Section \ref{HDRBG} discusses the DRBG algorithm. Section \ref{test} presents the test results of randomness test suites and in section \ref{Conclusion}, we have concluded our work.

\section{Enhanced-NRBG architecture}\label{architechture}

A NRBG \cite{nist90c} is a mechanism for producing random bits with full entropy. It has access to an entropy source and (when working properly) produces output bit strings that have full entropy. These bits are expected to be
indistinguishable (in practice) from an ideal random sequence to any adversary. It is also known as True Random Number (or Bit) Generator.
The NRBGs specified in  NIST SP 800-90C is referred to as enhanced-NRBGs since it includes a DRBG mechanism and a TRNG source. The advantages of using enhanced-NRBGs \cite{nist90c} are as follows:
\begin{enumerate}
\item  it provides a safe fall back mechanism i.e., provided  the DRBG   has been instantiated
securely, and the entropy source fails due to some reason, the  NRBG will then act as a DRBG.
\item  The DRBG mask if any adverse effect due to  even small deviation of entropy source from expected behaviour and reduces the damage.
\end{enumerate}

\begin{figure}
\centering{}\includegraphics[scale=0.35]{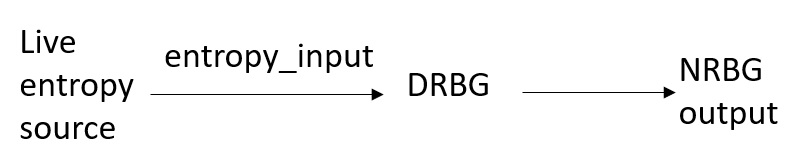}\caption{Block diagram of NRBG construction}\label{blockdia}
\end{figure}

In this work, we refer to the oversampling construction of enhanced-NRBG.  This NRBG is based on using a live entropy source that
provides entropy input for a constructed DRBG so that full entropy is provided. A live entropy source is an approved entropy source that can provide the requested amount of
entropy immediately or within an acceptable amount of time as determined by the user or
application requesting random bits from an RBG.
Conceptually, the DRBG mechanism accesses a 
live entropy source to obtain prediction
resistance and returns a number of bits equal to
half the security strength of the DRBG instantiation during a single NRBG request. This results
in full-entropy outputs and prediction resistance from the DRBG itself.
In Fig. \ref{blockdia}, we have presented the block diagram of NRBG construction.

\begin{figure}
\centering{}
\includegraphics[scale=0.35]{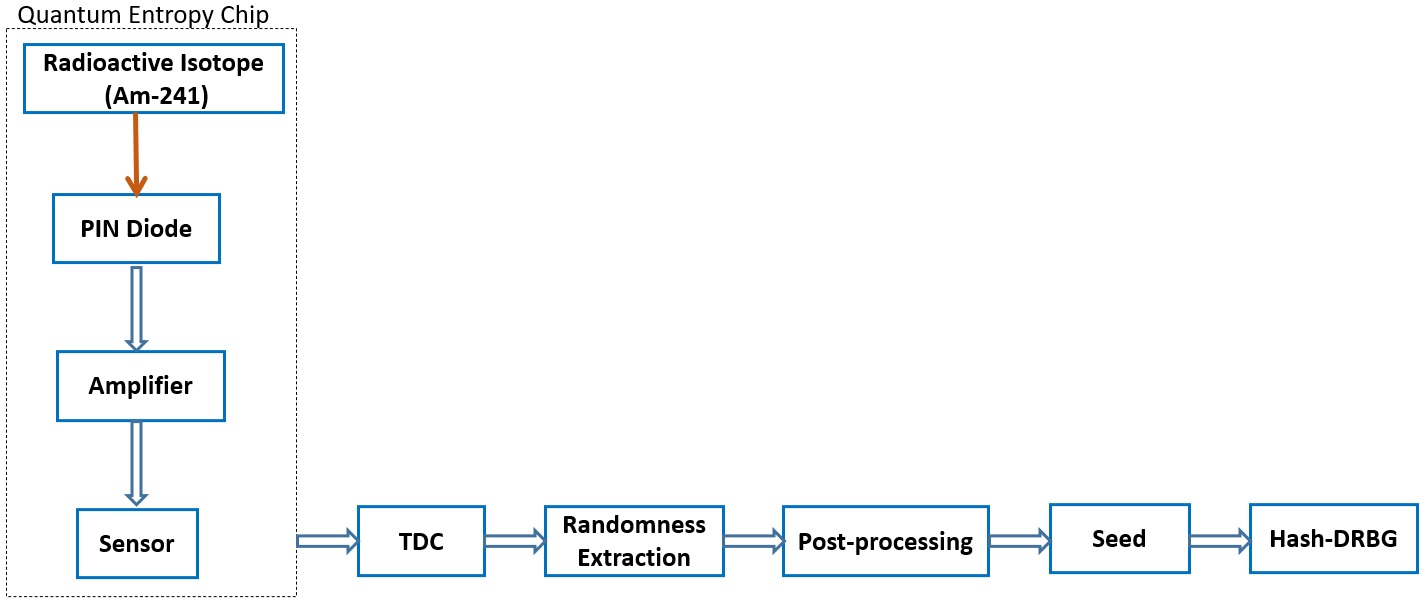}
\caption{Block diagram of the enhanced-NRBG}\label{ENRBG}
\end{figure}

The block diagram of our enhanced-NRBG is shown in Fig. \ref{ENRBG}. The set-up consists of a  quantum entropy chip (QEC) \cite{QEC}, a time-to-digital converter (TDC), a randomness extractor, a post-processing module and a randomness expansion module (Hash-DRBG). QEC exploits the radioactive decay of americium-241 to generate randomness. The emitted alpha particles from the decay are detected using a PIN diode. The generated decay impulses are amplified and digitized into time-stamps using TDC. Thereafter, a randomness extraction algorithm converts the TDC output into random digits. To remove the bias from the raw random numbers, post-processing is carried out using the Toeplitz hashing extractor, which results in compressed data with full entropy. A NIST-approved Hash-DRBG has access to this entropy and uses it as a seed during DRBG instantiation in order to generate its initial state or during reseeding.

\section{Quantum entropy source}\label{QEC}
The source of entropy  \cite{QEC}  is a quantum process, which is inherently
random and probabilistic, making it an ideal choice for generating unpredictable 
random numbers. The QEC  has radioactive decay as its source of randomness.
Radioactive decay (also known as nuclear decay, radioactivity, radioactive
disintegration or nuclear disintegration) is the process by which
an unstable atomic nucleus loses energy by spontaneous emission (without
any excitation from outside) of particles and radiation to form a
stable product. A material containing unstable nuclei is considered
radioactive. 
QEC exploits the emitted alpha particles resulting from the decay
of a radioactive isotope (Americium-241). The alpha particle is actually
the nucleus of a helium-4 atom $\prescript{4}{2}{\mathbf{He}}$,
consists of two protons and two neutrons, i.e. with two positive charges.
Alpha decay occurs when a heavy unstable (because of the excess of
nucleons) atomic nuclei dissipate excess energy by spontaneously ejecting
an alpha particle.
The alpha decay of Americium-241 can be expressed as:

\begin{equation}
\prescript{241}{95}{\mathbf{Am}} \rightarrow \prescript{237}{93}{\mathbf{Np}}+\prescript{4}{2}{\mathbf{He}}
\end{equation}

where, the numbers in subscript and superscript in the above reaction
represent the atomic number and the mass number of the nucleus, respectively.

QEC consists of a radioactive isotope (Am-241) that emits alpha particles
as a result of its decay, CMOS-type photo diode (for the detection of emitted alpha particles), two trans-impedance amplifiers (TIAs) (to amplify and detect low levels of the light current by the absorption of an alpha particle) and a comparator (to transform
the amplified voltage from TIAs to an analog pulse signal, called a quantum random pulse). The energy level of the alpha particle emitted by the Americium-241 used in the QEC is 4 MeV and its radioactivity level is 4.07 kBq. The chip has a size of 3 mm x 3 mm x 0.85 mm with a power consumption of 3 mW.  It generates random analog pulse when an alpha particle from the radioactive decay is detected which is then digitized and fed into time-to-digital converter. The output is processed by the randomness extraction module to generate raw random bits.

\subsection{Statistical modeling of the radioactive decay}

The probability of any given atom to decay in a time interval $(t,t+dt)$
is given by a negative exponential random variable and is expressed
as follows \cite{review}:

\begin{equation}
p(t)dt=\lambda e^{-\lambda t}dt
\end{equation}

where, $\lambda$ is the decay constant of the radioactive material.
Under the condition that, the amount of decaying atomic nuclei is
large enough to be considered as constant during the measurement time
and the half-life of the isotope is large enough so that the decay
constant $\lambda$ does not change, the time between consecutive
decay events is also an exponential random variable.

One significant aspect of the exponential distribution is its memory-less
property (also called Markov property). It states that the distribution
of the time interval between two successive event points is the same
as the distribution of the time interval between an arbitrarily chosen
point and the next event point. The time intervals are independent
of previous results, i.e. the decay pulses arrive at independent
times and the number of pulses that arrive in a fixed time period
follows a Poisson distribution. The probability of registering $k$
impulses within the interval $\Delta t$ is

\begin{equation}
p(k)=\frac{(\lambda\Delta t)^{k}}{k!}.e^{-\lambda\Delta t}.
\end{equation}

As mentioned earlier, QEC produces  random  pulses when these
emitted alpha particles are detected by the sensor. The length of
the time interval between two consecutive alpha decay pulses is
unpredictable and this fact can be utilized to generate true random
numbers.

\section{Randomness extraction}\label{algo}

\begin{figure}
\centering{}\includegraphics[scale=0.15]{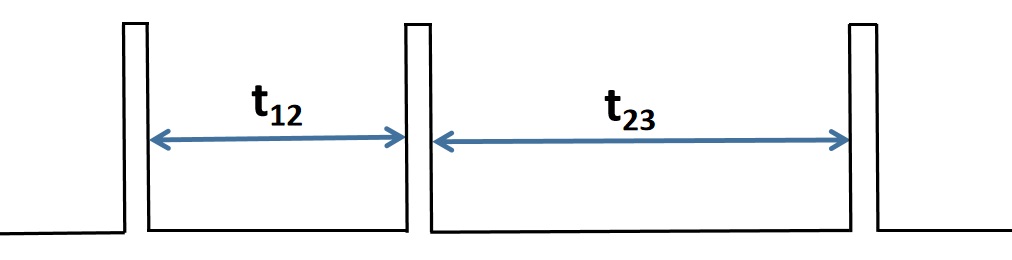}
\caption{Timing diagram}
\label{fig:interval}
\end{figure}

The randomness extraction from the radioactive decay  can be performed by multiple methods of randomness generation prescribed in \cite{randy}. In one method, say method 1,  we count the number of detected random events within a fixed time window. This method is based on the fact that the number of events that occur in a fixed time period is random and it can not be predetermined, since the radioactive $\alpha$-decay process follows a Poisson distribution. In another method, method 2, we consider the time interval between two consecutive pulses.  As a consequence of the memory less property, the time interval between successive decay pulses is a random variable. The random bits
here are obtained by deciding whether the time interval between two
pulses consists of an even or odd amount of timing units. The time
interval between two events is measured and is processed as follows:
a bit value of 1 is assigned if the interval length consists of an odd number of the time resolution units and a bit value of 0 is assigned if the interval length consists of an even number of the time resolution units. In yet another method, which is method 3, we measure the interval between consecutively detected decay events. The time between two consecutive pulses is stored as $t_{12}$ and compared to the time between the next two pulses $t_{23}$, as shown in Fig.  \ref{fig:interval}. If $t_{12}>t_{23}$, then a bit with value zero is generated otherwise, the bit generated will be one or vice versa. 
We collected data from 5 QEC in parallel for a sec and generated raw random numbers  from method 2 and 3. In Table \ref{tab:randomness methods}, we have compared the randomness of conditioned data from  method 2 and 3 in ENT test suite and we find that  their performance is comparable which is of no surprise. Hence, either of the methods are good candidates for generating random numbers.

\begin{table}
\centering{}
\caption{Randomness of conditioned data from method 2 and 3.}
\label{tab:randomness methods} 
\begin{tabular}{llll}
\hline\noalign{\smallskip}
ENT Test item &      Method 2 &   Method 3 &  Ideal value\\
\noalign{\smallskip}\hline\noalign{\smallskip}
Entropy (bits per byte) &    7.999998 &   7.999998 &   8.000000\\
Compression &     0\% &   0\%  &   0\% \\
Chi-square distribution &    29.36\% &   26.55 &   10\% $\sim$ 90\%\\
Arithmetic mean value &     127.5138 &   127.5027   & 127.5000\\
Monte Carlo value for Pi &    3.141224498 &   3.141678770   & 3.141592654\\
Serial correlation coefficient &     0.000081 & 0.000055  & 0.000000\\
\noalign{\smallskip}\hline
\end{tabular}
\end{table}

We have adopted  method 3 and generated n bits from n+1 decay events. The value of n is on an average of 45 bits per sec. Interestingly, 
the web-based random number server HotBits \cite{hotbits,Aleksandr}
is based on similar principles. Finally, the outputs from 15 QECs are concatenated to provide an input of at least 600 bits of entropy.  The min-entropy of the data provided is on an average of  0.8.

\section{Post-processing using Toeplitz hashing extractor}\label{PP}

The random numbers generated post digitization will have biases due to the exponential waiting time between two events and it will be mixed with classical noise particularly from manufacturing defects,  electronic noise and design methodology of the entire system.   Hence, the resultant bits undergo post-processing or conditioning for removing the bias. The compressed bits are unbiased and unpredictable. Toeplitz hashing is proven to be information-theoretic secure \cite{pp} and prescribed for quantum key distribution and QRNG.

Toeplitz hashing is performed by the multiplication of a Toeplitz matrix with the raw random bits obtained post digitization. The Toeplitz matrix is a matrix where the diagonal elements are similar. To produce such a matrix, we only need elements of the first row and the first column. Thereafter, we can generate the whole matrix. The first column elements are chosen randomly using an RNG, in our case, we have used a home-built TRNG. With this column as a seed to a linear feedback shift register, the first row is generated. A primitive polynomial is also used in the generation of the first row elements. The size of the matrix is taken as $476 \times 600$. We have calculated the average min-entropy of the raw random data, which comes out to be 0.8. The chosen dimensions of the matrix compress the data by $20\%$. The data is then converted into hexadecimal numbers and given to the Hash-DRBG.

\section{Implementation of Hash-DRBG}\label{HDRBG}

Hash-Deterministic Random Bit Generator (DRBG) produces
random numbers through computations by making use of cryptographic hash functions approved by NIST. We have implemented the SHA-256 hash function in the DRBG for producing random data. When the entropy input given to the DRBG has sufficient randomness, the bits produced through the DRBG will also be random with a high probability in the absence of any pattern.  This section explains the working of the Hash-DRBG. An open-source GitHub repository is used as a reference for the implementation \cite{python}.

The NIST standard  \cite{nist90a} claims that each of the NIST DRBGs (including
Hash-DRBG) is backtracking resistant
and prediction resistant. The former
security property is the familiar forward security notion for PRNGs,
which guarantees that in the event of a state compromise, the output produced
prior to the point of compromise remains secure. The latter property
ensures that if the state is compromised and subsequently reseeded
with sufficient entropy, then security will be recovered.

Toeplitz hashing compresses the initial entropy of around 600 bits
to 476 bits. The entropy block size is chosen based on the available
polynomial \cite{Zierler}. These 476 bits are then converted to hexadecimal
values by converting every four bits into one hexadecimal number.
This string of hexadecimal numbers is the seed material that produces
the seed and other internal states for the DRBG. The DRBG contains
three main functions namely, Instantiate, Generate and Reseed which
are used for producing the random data \cite{nist90a}. The program produces
1 Gb of data by placing 2000 requests. The requests are handled in
parallel using the multiprocessing library in python. For each request,
the DRBG produces 5,00,000 bits. The produced bits are concatenated
to form a stream of 1 Gb of random bits. The program takes around 3
to 5 seconds to produce 1 Gb of random data using 6 processors running
in parallel. NIST suggests that each request can produce a maximum of
$2^{19}$ bits. Therefore, each request made in the implemented DRBG
only generates 5,00,000 bits $<$  ( $2^{19}$). NIST also recommends that
only a maximum of $2^{24}$ requests to be placed for a single seed
after which the DRBG should be provided with a different seed (reseeding).
Since only 2000 requests are needed for generating 1 Gb of data, it
is assumed that no reseeding is required for producing 1 Gb of random
bits. The standard says that the nonce, personalization string and
additional input are optional and their absence does not reduce the
security of the system. Therefore, they are not explicitly provided
but the input of 476 bits is more than the required minimum length
of entropy. It is assumed that the concatenation of each set of 5,00,000
bits in any order do not affect the randomness of the produced bits. As SHA-256 is used for hashing, the security strength of the Hash-DRBG is at least 256 bits. This means, for a random number generator of security strength 256 bits, $2^{256}$ operations should be done to figure out the seed.

\begin{figure}[h]
\centering{}\includegraphics[scale=0.45]{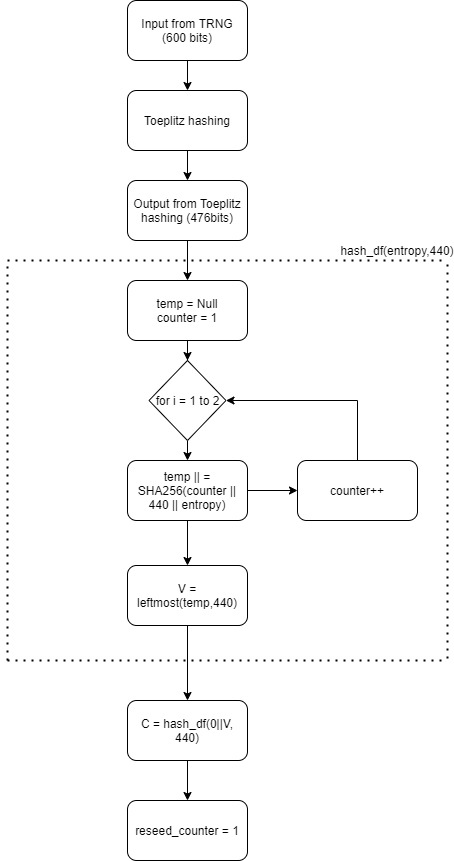}
\caption{Flowchart showing Instantiation.}
\label{fig:ins}
\end{figure}

The flowchart of DRBG instantiation is shown in Fig. \ref{fig:ins}. The DRBG algorithm uses an instantiate function for setting the internal values of the DRBG. These internal values are seed, a constant and a counter called reseed counter. The seed is used for generating the random bits. It is initially constructed by hashing the entropy input and changes whenever the Generate function is called for producing random bits. It is also modified when the Reseed function is requested. The NIST standard suggests that the seed should be of length 440 bits if SHA-256 is used for hashing. A constant is created by hashing the seed. This constant is used for modifying the seed after the generation of random bits. This constant is set during the instantiation of DRBG and changes whenever the Reseed function is requested. This is also of length 440 bits as prescribed by NIST. The reseed counter keeps track of the number of times the Generate function is called. It is incremented whenever a set of random bits are produced by the Generate function. The maximum number of requests to the Generate function that can be made is called the reseed interval. When the reseed counter exceeds the reseed interval, Hash DRBG should be reseeded. The hash derivation function (hash\_df) is used to produce the seed and the constant. It hashes the input repeatedly with some modifications and appends it to an empty string. This string of hashed values is used as the seed or the constant.

\begin{figure}
\centering{}\includegraphics[scale=0.45]{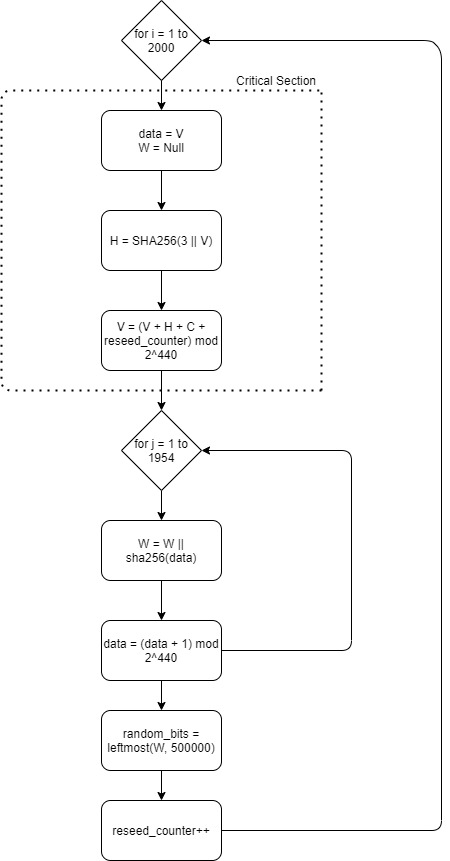}
\caption{Flowchart showing Generation of bits.}
\label{fig:gen}
\end{figure}

The flowchart of the Generate function is shown in Fig. \ref{fig:gen}. The random bits are generated through the generate function. As the output of SHA-256 is of length 256 bits, repeated hashing of the seed is done for producing the desired number of random bits. After each hashing, the input to the hash is incremented in addition modulo $2^{440}$. The hashed bits are appended in a null string and returned as a single string of random data of the requested length. After the random bits are produced, the seed is modified. The seed is hashed and the hashed value is used for the modification of the seed. The addition (in modulo $2^{440}$) of the hashed value of the seed, the seed, the constant and the reseed counter value is used to get the modified seed. This modified seed is then used for the next request to the generate function. Since only 500000 bits can be produced in a single request, the function is executed 2000 times in parallel to produce 1 Gb of random data. Since the seed(V) is used by each instance of the Generate function, the modification of the seed is inside the critical section. 

To summarize, the raw bits are read from a text file and stored in a list. This list of bits is then given to the Toeplitz hashing function. The Toeplitz hashing function initialises the LFSR with random initial state. The initial state vector is used as the first column of the Toeplitz matrix. The chosen polynomial is used for producing the further states. These states are used as the first row elements of the Toeplitz matrix. With the Toeplitz matrix fully generated, privacy amplification of the raw bits is done by multiplication of Toeplitz matrix with the data. These processed entropy bits are then converted into hexadecimal format. The hexadecimal numbers are used to create an instance of the Hash DRBG class. The instantiate function sets the value of seed, constant and reseed counter. A for-loop is run for 2000 iterations. Each iteration calls the Generate function to produce 1,25,000 hexadecimal numbers which are then converted to 5,00,000 bits. This for-loop is executed in parallel using the 6 cores available in the system. Combining all the sets of 5,00,000 bits gives 1 Gb of random data. We have used the LFSR function in the Pylfsr module is used as the Linear Feedback Shift Register whose states are used as the first row elements of the Toeplitz matrix. Scipy version 1.6.0 is used for the multiplication of Toeplitz matrix with the raw bits. Numpy is used for matrix operations and Pandas is used for excel handling. The multiprocessing module in python is used for parallel programming. The code is executed in AMD Ryzen 5 4500U hexacore processor with base clock @2.38GHz.

\section{Randomness evaluation and statistical analysis}\label{test}

It is crucial to evaluate the strength of randomness of any random number generator and this is achieved using statistical test suits. Most randomness tests check one or more
statistical properties of long sequences of random numbers. We have
evaluated the final random bits generated from the Hash-DRBG with different measures of randomness, for example, frequency distribution, auto-correlation and statistical test suite 
NIST  \cite{NISTtest,ent}. We evaluated 5 sets of 1 Gb data and the tested files passed all the above-mentioned tests. 

\subsection{Auto-correlation}

\begin{figure}
\centering{}\includegraphics[scale=0.65]{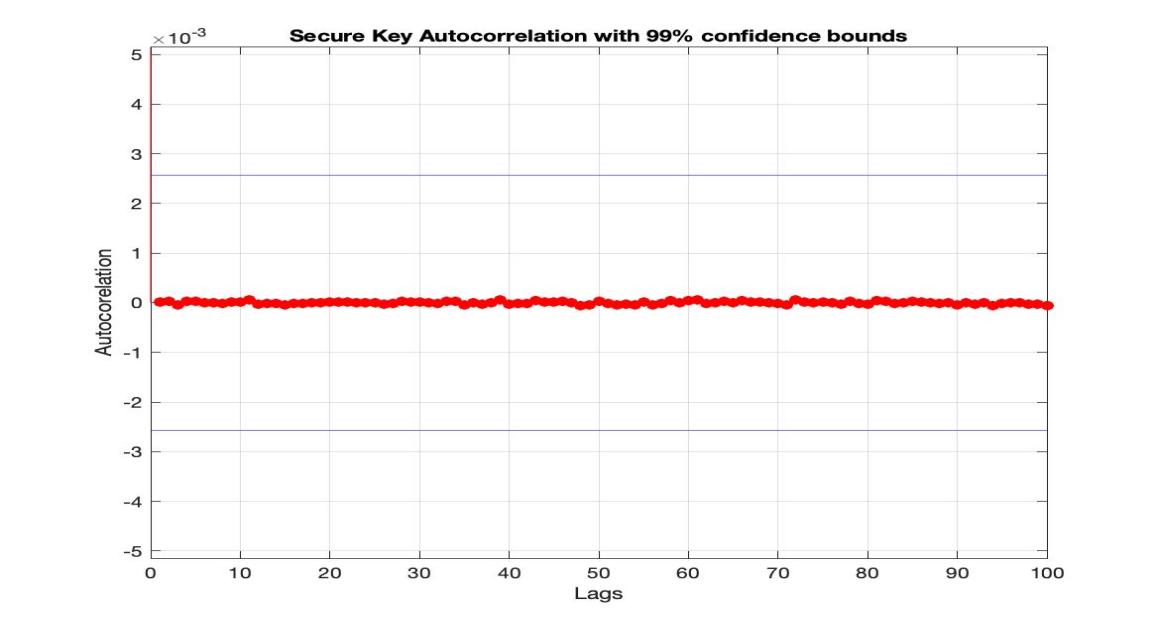}
\caption{Auto-correlation plot}
\label{fig:Auto-correlation}
\end{figure}

Auto-correlation checks for correlations among bits in a data
set at varying lags between the bits in the sequence. If random, auto-correlations should be near zero for all lag separations. Fig. \ref{fig:Auto-correlation}
shows the auto-correlation plot for bitstream generated for lag 0
to 100. The results indicate the absence of significant correlations in generated bitstreams.

\subsection{Uniform distribution of random number data}

\begin{figure}[h]
\centering{}\includegraphics[scale=0.30]{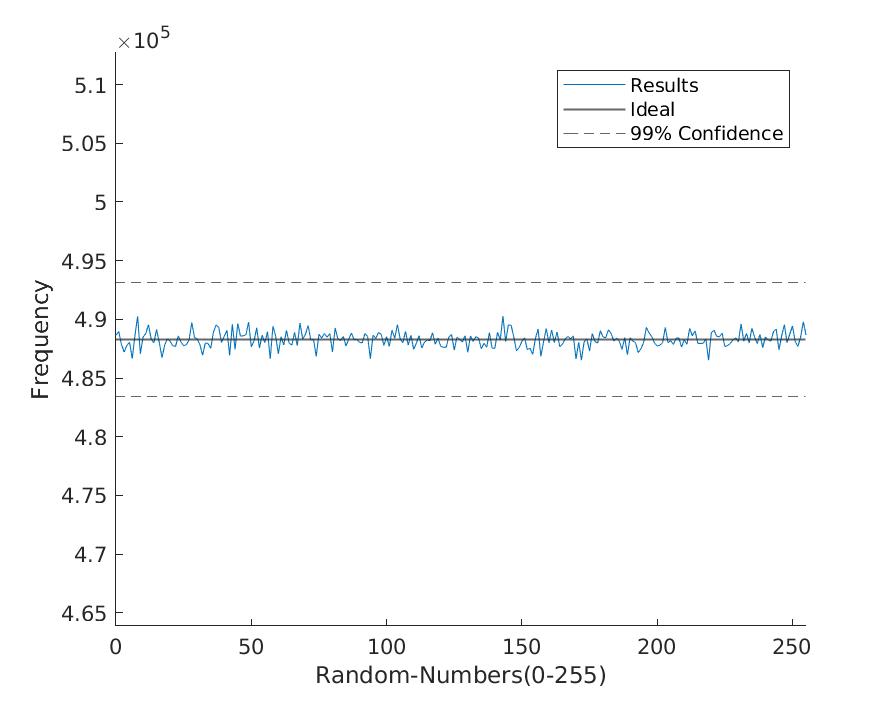}
\caption{Frequency of occurrence of each random number}
\label{fig:distribution}
\end{figure}

We collected 1.25 Gb of random data. We converted consecutive 8 bits to a byte in the range of  0 to 255. Fig. \ref{fig:distribution}, depicts the distribution of the frequency of values across the range 0-255. For a perfectly random sequence, the frequencies are uniformly distributed. As can be seen in the plot, the collected data gives the consistent distribution of frequencies, indicating that there is no tendency for specific numbers to occur more or less, hence a uniform probability distribution is generated.

\subsection{NIST }

\begin{figure}
\centering{}
\includegraphics[scale=0.5]{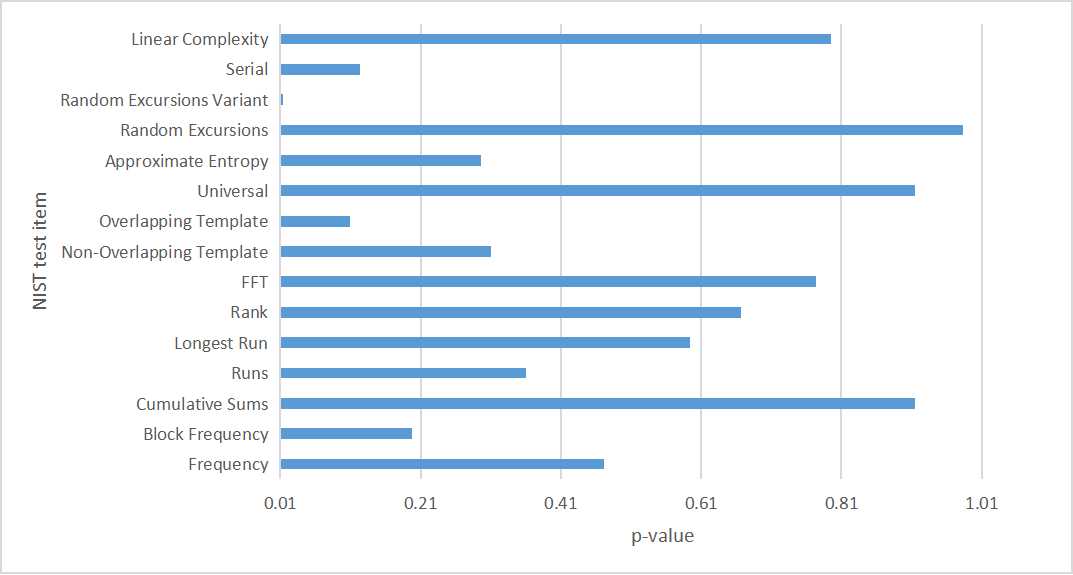}
\caption{p-value plot for the NIST tests for a typical
sequence of 1.25 Gb.}
\label{fig:p-value-plot}
\end{figure}

\begin{figure}
\centering{}
\includegraphics[scale=0.5]{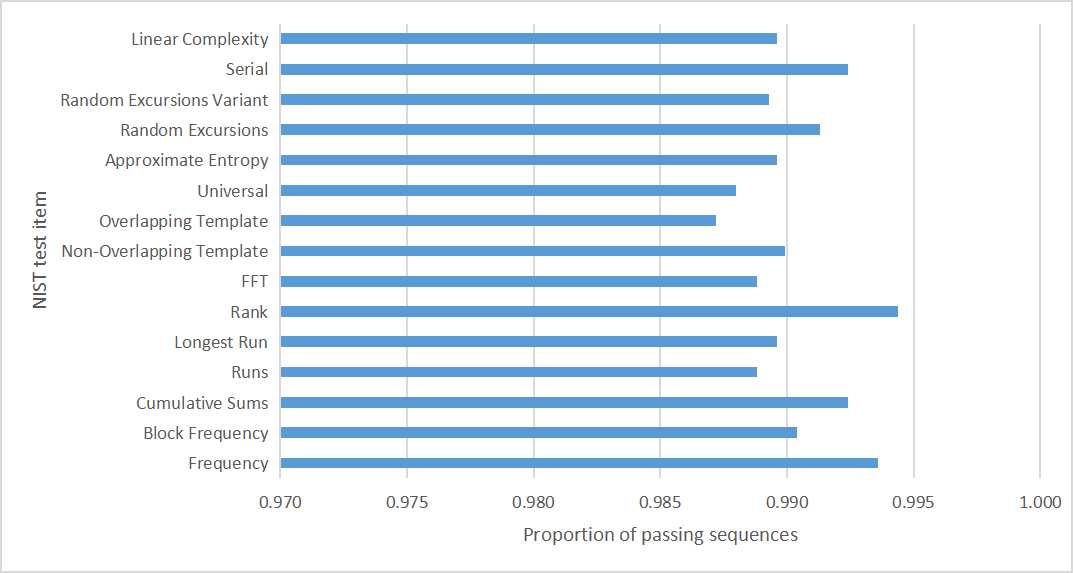}
\caption{Confidence plot for the NIST tests for
a typical sequence of 1.25 Gb.}
\label{fig:Confidence-plot}
\end{figure}

NIST 800-22 \cite{NIST} proposes a statistical package consisting of  15 measures of randomness. It computes a p-value for each test which should be larger than the significance level for the
test to be considered successful.  There are certain tests that produce multiple outcomes of p-values
and proportions, Fisher's method is used to combine the p-values to
form a final p-value.
The reason for using Fisher's method instead of taking the worst case
result or average is based on a basic axiom of probability: p-values
are probabilities and probabilities for the outcomes of independent
tests don't add, they multiply. Fisher's method is based on the product, since we multiply independent probabilities to find their joint probability. As seen in Fig. \ref{fig:p-value-plot} and \ref{fig:Confidence-plot}, the p-values are routinely above the significance level and the proportions
are within the bound, hence all the 15 tests in the NIST statistical
test suite are passed at the $\alpha$ = 0.01 significance level.

\begin{table}
\centering{}
\caption{Results of ENT tests for a typical sequence
of 1 Gb.}
\label{tab:Results-of-ENT}
\begin{tabular}{lll}
\hline\noalign{\smallskip}
ENT Test item &   Enhanced-NRBG &   Ideal value\\
\noalign{\smallskip}\hline\noalign{\smallskip}
Entropy (bits per byte) &   7.999999 &   8.000000\\
Compression &   0\% &   0\% \\
Chi-square distribution &   67.83\% &   10\% $\sim$ 90\%\\
Arithmetic mean value &  127.5020   & 127.5000\\
Monte Carlo value for Pi &   3.141046706   & 3.141592654\\
Serial correlation coefficient &  -0.000093   & 0.000000\\
\noalign{\smallskip}\hline
\end{tabular}
\end{table}
\subsection{ENT }

ENT test suite contains the following statistical tests: entropy,
chi-square, arithmetic mean, Monte Carlo and serial correlation coefficient tests. As evident from the results reported in Table \ref{tab:Results-of-ENT}, the generated random numbers passes all the tests in the ENT test suite.

\section{Conclusion}\label{Conclusion}

We have designed and demonstrated an enhanced-NRBG that delivers high-quality, high-speed random numbers. We have discussed the algorithm for the generation of random numbers from QEC and presented a detailed implementation of Hash-DRBG.  Hash-DRBG  receives its seed from a live quantum entropy source, which is a series of  radioactivity-based quantum entropy chip. Finally, we achieved a throughput of 1 Gb and the  random bit sequence  passed the  standard statistical test suites of ENT and NIST. Also, the uniform distribution and auto-correlation analysis confirm the quality and reliability of the random bits produced. 
\bibliographystyle{unsrtnat}
\bibliography{references}

\end{document}